\documentclass[11pt,eqs]{article}

\usepackage{latexsym,amsmath}
\usepackage{mathtools}
\usepackage{amssymb,tikz}
\usepackage{xcolor}

\usepackage[margin=1in]{geometry}

\usepackage{latexsym,amsmath,amssymb,amsthm,amstext}
\usepackage{fancyhdr}

\usepackage{latexsym}
\usepackage{amsmath}
\usepackage{graphicx,relsize}
\usepackage{amstext,amssymb}
\usepackage{amsmath,tikz}
\usepackage{mathtools}

\usepackage{multirow,multicol}
\usepackage{pdflscape}

\usepackage{latexsym,amsmath}
\usepackage{mathtools}
\usepackage{amssymb,tikz}
\usepackage{xcolor}

\usepackage[margin=1in]{geometry}

\usepackage{latexsym,amsmath,amssymb,amsthm,amstext}
\usepackage{fancyhdr}

\usepackage{latexsym}
\usepackage{amsmath}
\usepackage{graphicx,relsize}
\usepackage{amstext,amssymb}
\usepackage{amsmath,tikz}
\usepackage{mathtools}

\usepackage{multirow,multicol}
\usepackage{pdflscape}

\def\cleq{\setcounter{equation}{0}}
\title{Brackets in bosonic string theory
\thanks{Work supported in part by Institute of Physics, University of Belgrade, Serbia.}}
\author{ Lj. Davidovi\'c \thanks{e-mail: ljubica@ipb.ac.rs}\\
{\it Institute of Physics, University of Belgrade}\\
{\it Pregrevica 118, 11080 Belgrade, Serbia}}

\begin{document}
\maketitle

\begin{abstract}

We give a review of brackets and interior products in bosonic string theory,
in different representations, used in formulation of a theory
and derived in a transformation of related mathematical structures.
We consider the $C$-bracket, defined in double string theory,
in a context of Poisson algebra of symmetry generator,
and show  its connection to prior brackets trough anomaly reduction,
and its characteristics related to Courant algebroid definitions.

\end{abstract}

\section{Introduction}

The point particle can be described using Lagrangian or Hamiltonian formalisms, the first
given in a configuration space $\{x^\mu\}$ and the second in the phase space $\{x^\mu,\pi_\mu\}$. Given a system of {\it differential
equations} describing {\it  kinematic properties} of a particle in the configuration space, one can define
a Lie derivative of the arbitrary function of coordinates along the solution, as a time derivative
of that function
\begin{equation}
\frac{df}{dt}=\frac{\partial f}{\partial x^{i}}\frac{\partial x^{i}}{\partial t}=
v_{i}\frac{\partial }{\partial x^{i}}f={\cal L}_{v}f.
\end{equation}
It is given in terms of vector fields (elements of the tangent vector space), acting on themselves by a Lie bracket.
In a phase space the time evolution of a function of phase space variables is governed by the Hamiltonian ${\cal H}$ and
Poisson bracket
\begin{equation}
\frac{df}{dt}=\{f,{\cal H}\}.
\end{equation}
 The same as the canonically conjugated variables, coordinates and momenta, are related by the Poisson bracket,
elements of tangent space are related to elements of the cotangent space by a natural parring (contraction):
\begin{eqnarray}
&&\{x^\mu,x^\nu\}=0,  
\nonumber\\
&&\{x^\mu,\pi_\nu\}=\delta^\mu_\nu,\qquad \langle dx^\mu,\frac{\!\!\!\!\partial}{\partial x^\nu}\rangle=\delta^\mu_\nu.
\nonumber\\
&&\{\pi_\mu,\pi_\nu\}=0,  
\end{eqnarray}
Trough parring the Lie derivative action on differential forms is found.
The same as the functions are Taylor expanded into power series of canonical variables, differential forms are expanded into order-wise power series of elements of a cotangent space.
Order defines the orientation of orts on the spanned surfaces.

Vector fields form a 
modulo over the ring of smooth functions, and their dual space is of linear functionals, differential forms, while the connection between spaces is realized by differentials and inner products. In this bimodulo the standard differential operators: divergence and curl are seen to be easily calculated for the volume form, the first trough the inner product with a vector field
$\vec{v}=v_{x}\frac{\partial\;}{\partial x}+v_{y}\frac{\partial\;}{\partial y}+v_{z}\frac{\partial\;}{\partial z}$
 and the second trough differential of its corresponding form
$\sigma(\vec{v})=v_{x}dx+v_{y}dy+v_{z}dz$ \cite{Bolibruh}.

In hamiltonian dynamics, one introduces the Hamiltonian vector field, the first order differential operator, $X_{f}=\sum_{i}\Big{(}
\frac{\partial f}{\partial x^{i}}\frac{\partial }{\partial \pi_{i}}-
\frac{\partial f}{\partial \pi_{i}}\frac{\partial }{\partial x^{i}}
\Big{)}$,
through which the Jacobiator of Poisson bracket is given by
\begin{equation}
\{f,\{g,h\}\}+cycle=[X_{f},X_{g}]h-X_{\{f,g\}}h,
\end{equation}
where bracket $[,]$ marks the commutator. 
Linear functions on phase space, with Poisson bracket form a Lie algebra, because
$\{f,{\alpha g}\}=\alpha\{f,{ g}\}$, $\{f,{g}\}=-\{g,{ f}\}$ and the Jacobi identity is satisfied.
In differential geometry, the Poisson bracket is generalized to
$\{f(x),g(x)\}=\Pi_{ab}\frac{\partial f}{\partial x^a}\frac{\partial g}{\partial x^b}$ with $\Pi_{ab}=-\Pi_{ba}$,
and its Jacobi identity is equivalent to annihilation of the following structure
\begin{equation}
\Pi^{ao}\partial_{o}\Pi^{bc}+\Pi^{bo}\partial_{o}\Pi^{ca}+\Pi^{co}\partial_{o}\Pi^{ab}.
\end{equation}
The Jacobi identity is itself then equivalent to Leibniz rule for a derivative defined by $d_{f}\,\cdot\equiv \{f,\cdot\}$:
$d_{f}\{g,h\}=\{d_{f}g,h\}+\{g,d_{f}h\}$.
Manifold with this kind of Poisson bracket (or Poisson bivector) is called Poisson manifold \cite{CF}.

If one introduces the order-wise product in the tangent space as well and considers $\Pi=\frac{1}{2}\Pi^{ab}
\frac{\partial\;\,}{\partial x^{a}}\wedge\frac{\partial\;\,}{\partial x^{b}}$ then the Jacobi identity for the Poisson bracket
$\{f,g\}=\Pi(df,dg)$ is equivalent to vanishing 
of the Schouten-Nijenhuis bracket $[\Pi,\Pi]_{S}=0$ \cite{P}.  
$d$ is an exterior derivative connecting smooth functions to 1-forms, 1-forms to 2-forms, and so forth. It is a linear transformation, with $d^{2}=0$, acting between forms of orders differing by one, whose definition incudes the bracket relevant for transformation of vector fields.
A Poisson structure of this form
induces the Koszul bracket \cite{Kozdef} for differential forms defined by
\begin{equation}\label{eq:kozbra}
 [\alpha,\beta]_\pi={\cal L}_{\tilde\pi\alpha}\beta
-{\cal L}_{\tilde\pi\beta}\alpha
-d(\pi(\alpha,\beta)).
\end{equation}

The motion in gravity \cite{GR} on the arbitrary manifold can not be described using one coordinate system.
In coordinate systems overlaps, the transition from one to another is governed by diffeomorphisms.
The notions of locally and globally defined quantities were established,
with globally defined quantities,  called tensors, being the ones
transforming properly under these transformations.
Their own transformation is established in terms of connections, which define the parallel transport of vector fields and forms.
These are viewed as the additional structures on a manifold.

When a point particle is replaced by a notion of string \cite{GSW}, both Lagrangian
and Hamiltonian are defined on a world-sheet swapped by a string moving in the given background.
The world-sheet is parametrized by time-like and space-like parameters. The background fields obey the space-time equations of motion, arising from the {\it Virasoro algebra} for the
{\it energy-momentum tensor}. The similarity transformations 
\begin{equation}
T_\pm\rightarrow e^{-i\Gamma}T_\pm e^{i\Gamma}
\end{equation}
do not break the algebra
and induce the symmetry transformations for background fields \cite{EO}. 
The Lie algebra of these transformations 
was found to define
a number of new  brackets in string theory governing the composition of symmetry parameters.
It was shown that the same appear in Poisson algebra of
the generators of string`s background symmetries and current algebra \cite{Alekseev}.

The dynamics of a string is described by a hamiltonian, standardly used in couple of different forms:  except for a defining form
\begin{equation}
{\cal H}= \pi_\mu \dot{x}^\mu - \mathcal{L} = \frac{1}{2 \kappa} \pi_\mu (G^{-1})^{\mu \nu} \pi_\nu - 2 x^{\prime \mu} B_{\mu \nu} (G^{-1})^{\nu \rho} \pi_\rho + \frac{\kappa}{2} x^{\prime \mu} G^E_{\mu \nu} x^{\prime \nu},
\end{equation}
where $G^E_{\mu \nu} = G_{\mu \nu} - 4 (B G^{-1} B)_{\mu \nu}$ is the open string effective metric;
expressed in terms of currents, components of energy-momentum tensor
\begin{equation}
{\cal H}= \frac{1}{4\kappa}(G^{-1})^{\mu\nu}\big{(}
j_{+\mu}j_{+\nu}
+j_{-\mu}j_{-\nu}
\big{)},\quad
j_{\pm\mu}=\pi_\mu+2\kappa\Pi_{\pm\mu\nu}x^{\prime\nu},
\end{equation}
and
the most often, in its  matrix representation \cite{MV,SIEG1}
\begin{equation}
{\cal H}=\frac{1}{2\kappa} (X^T)^M H_{MN} X^N,\qquad 
X^M = \begin{bmatrix}
\kappa x^{\prime \mu} \\
\pi_\mu \\
\end{bmatrix},
\end{equation}
given in terms of a generalized metric 
\begin{equation}
H_{MN} = 
\begin{bmatrix}
G^E_{\mu \nu} & - 2B_{\mu \rho} (G^{-1})^{\rho \nu} \\
2(G^{-1})^{\mu \rho} B_{\rho \nu} & (G^{-1})^{\mu \nu}
\end{bmatrix}.
\end{equation}
In the last case, sigma model hamiltonian is expressed through the generalized phase space coordinate, with generalized metric depending on initial background fields and an effective metric.

In string theory a T-duality was discovered \cite{GPR}, connecting different but physically equivalent theories,
and the procedures to find equivalent actions were established \cite{B}.
T-duality opened the search for the connections between
the relevant mathematical structures within T-dual theories, 
in its basic variant it interchanges coordinate spatial derivatives and momenta.
It connects background fields of the closed string to the notions known to the open string:
effective metric (metric obtained for the solution of its boundary conditions) and
form of the noncommutativity of coordinates on its end-points.
The noncommutativity parameter for coordinates measures their nonasociativity.
T-duality connects theories with different commutativity properties and
transforms bosonic string symmetries into one another.

The string theory is defined on a world-sheet, a two dimensional object swapped by a string,
moving under the influence of space-time background.
Introducing a notion of projection
one can interiorize any function as a layer (section).
For instance $f:\mathbb{R}^{4}\rightarrow \mathbb{R}^{2}$  is usually represented as a graph: set of points $s=(x,f(x))$ in $\mathbb{R}^{4}\times \mathbb{R}^{2}$.
However it can also be understood as a layer because by a
projection to the first one can reassemble the initial function:
it is a layer defined by $\pi\circ s=id$ \cite{Bolibruh}. 
On the world-sheet,
within considered descriptions, one affiliates  the following sets of variables
\begin{equation}
\frac{\partial\;}{\partial x^\mu}\,;\;dx^\mu
\quad\longleftrightarrow\quad
\pi_\mu\,;\;\kappa x^{\prime\mu}
\quad\longleftrightarrow\quad
\pi_\mu\,;\;\pi^{\star\mu}
\end{equation}
representing tangent $TM$ and cotangent $T^\ast M$; tangents on a world-sheet $\frac{\partial}{\partial\tau}x^\mu,\frac{\partial}{\partial\sigma}x^\mu\,$ first of them being Legendre transformed, and the second multiplied by a constant $\kappa$ for later convenience; and momenta of two T-dual theories.
While in mathematics first two both go to a point when the surface is shrinking to a point, in physics the calculus on hidden dimensions should  
provide the existing laws on nature, in geometry. 
Second and third sets are connected by basic T-duality:
T-duality projects the third pair to the second, when the T-dual theory is 
considered on the  phase space of the initial theory. 
Except for being a exterior derivative, $d$ in $dx^\mu$ can be a measure defined step over the equation of motion.
The connection between the first and the second pair
is also established, 
within  relations between non-basic constituents of a theory,
represented as points at  diagrams \cite{JV}
\begin{center}
\begin{tikzpicture}
\draw (1,0.25) node {$E$};
\draw[->] (1,0) -- (1,-1);
\draw (1.25,-0.5) node {$^{g_{E}}$};
\draw (1,-1.25) node {$\;\,E^\ast.$};
\draw[dashed][->] (-0.35,0.25) -- (0.75,0.25);
\draw(0.2,0.45) node {$^{\cal D}$};
\draw[->] (-0.65,-1.25) -- (0.75,-1.25);
\draw(0.05,-1.05) node {$^{\rho^{T}}$};
\draw[->] (-1.1,0) -- (-1.1,-1);
\draw (-1.15,0.25) node {$C^\infty(M)$};
\draw (-1.15,-1.25) node {$T^\ast M$};
\draw (-0.95,-0.5) node {$^{d}$};
\end{tikzpicture}
\end{center}
Authors in \cite{KS} define generalized differentials on $E$
demanding the conservation of natural parring
\begin{equation}
\langle{^{\scriptscriptstyle E}}\!df,\psi\rangle:=\langle df,\rho(\psi)\rangle=\rho(\psi)f,
\end{equation}
and
\begin{equation}
\langle{^{\scriptscriptstyle E}}\!d\omega,\psi\otimes\psi^\prime\rangle:=
d\omega(\psi,\psi^\prime)
=
\rho(\psi)\langle\omega,\psi^\prime\rangle
-\rho(\psi^\prime)\langle\omega,\psi\rangle
-\langle\omega,[\psi,\psi^\prime]\rangle,
\end{equation}
which for the
standard Lie algebroid $(TM, Id)$ itself becomes the definition of de Rham differential.

Because the string can wrap around compact dimensions, the notion of winding number is one of fundamentals in string theory.
T-duality was found to be connected to that fact, 
 it interchanges windings and momenta, so the consideration of 
both initial theory and its T-dual together lead to
double string field theory formulation \cite{T}, in which
the theory is defined in doubled $\{x^\mu,y_\mu\}$ configuration space \cite{D},
formulation which manifestly admits T-duality.

The generalized metric of DFT can be written in terms
of a Schur decomposition \cite{RS}. Half of the degrees of freedom are nonphysical and are 
excluded by imposing the strong constraint.
Strong constraint imposes that all products of fields and gauge parameters should annihilate under $\partial^{M}\partial_{M}$,
and it comes from the week constraint originating from the level matching condition, valid for any field or parameter \cite{HLZ}.
 Different polarisations induce
different T-duality frames which are related by an $O(d,d)$-transformations.

By forming a doubled space,
 one leaves the 
notion of geometric space and introduces the notion of nongeometric space (described by two sets of coordinates, mutually connected by T-duality),
the term nongeometricity was first used to underline the existence of coordinate noncommutativity.
Double coordinate naturally appears in  string theory T-dualizations,  as an argument of the T-dual fields. Regardless of weather the T-dualization is performed for all or some selected
directions, the T-dual field's argument always contains the whole double coordinate \cite{DNS}. 
Also, in the open effective string theory obtained for the solution of boundary conditions,
the argument of the effective fields is a double coordinate for the effective coordinates \cite{EPJC.3}. For effective theories the Poisson bracket is 
replaced with the Dirac bracket, which governs the evolution of constrained systems.

Two T-dual theories can be represented on the unique phase space in a symplectic T-duality:
\begin{center}
\resizebox{0.55\textwidth}{!}{
\begin{tikzpicture}
\draw[-] (-4.75,0.5) -- (-0.2,0.5);
\draw[-] (-5.35,0.57) -- (0.35,0.57);
\draw[-] (-4.75,0.5) -- (-4.75,-1);
\draw[-] (-0.2,0.5) -- (-0.2,-1);
\draw[-] (-3.7,-0.25) -- (-3.2,-0.25) -- (-3.2,0.3) -- (-3.7,0.3) --cycle;
\draw (-3.45,0) node {$i_\mu$};
\draw[-] (-1,-0.25) -- (-0.5,-0.25) -- (-0.5,0.3) -- (-1,0.3) --cycle;
\draw (-0.75,0) node {$\,{^\star\! i}^\mu$};
\draw[-] (-3.7,-1) -- (-3.2,-1) -- (-3.2,-0.45) -- (-3.7,-0.45) --cycle;
\draw (-3.45,-0.75) node {$k^\mu$};
\draw (-4.2,0) node {$j_{\pm\mu}$};
\draw (-1.5,0) node {$^\star j_{\pm\mu}$};
\draw (-4.2,-0.75) node {$l_\pm^\mu$};
\draw (-4.2,0.75) node {${\cal H}$};
\draw (-1.5,0.75) node {$^\star{\cal H}$};
\draw (-5.25,-0.75) node {$G_{E}$};
\draw (-5.25,0) node {$G^{-1}$};
\draw (0.25,0) node {$G_{E}$};
\draw[<->] (-3,-0.75) -- (-2,0.05);
\draw (-2.25,-0.55)  node {${\cal T}$};
\end{tikzpicture}}\\
\end{center}
Hamiltonian densities are built out of pair of currents and a  metric inverse,
which are not their own T-duals. So, the T-duality transforms the T-dual hamiltonian 
into the same but differently represented original hamiltonian.
The currents of two representations
are used to extract two pairs of new phase space variables
\begin{center}
\resizebox{0.75\textwidth}{!}{
\begin{tikzpicture}
\draw (-4.2,0) node {$j_{\pm\mu}$};
\draw[<->] (-3.7,0) -- (-2.6,0);
\draw (-3.2,0.2) node {$^{A}$};
\draw (-2.2,0) node {$^\star j_{\pm\mu}$};
\draw[<->] (-4.2,-0.4) -- (-4.2,-1.5);
\draw (-4.5,0-1) node {$^{T_{S}}$};
\draw (-4.2,-2) node {$l_\pm^\mu$};
\draw[<->] (-3.8,-2) -- (-2.3,-0.4);
\draw (-2.8,-1.4) node {$^{T}$};
\draw (-1.5,-1) node {$=$};
\draw (2.5,-1) node {$\pm\;\sim$};
\draw (-0.2,0) node {$i_\mu$};
\draw (3.8,0) node {$x^{\prime\mu}$};
\draw[<->] (0.5,0) -- (1.5,0);
\draw[<->] (4.5,0) -- (5.5,0);
\draw (1,0.2) node {$^{A}$};
\draw (5,0.2) node {$^{A}$};
\draw (2.2,0) node {$\,{^\star i}^\mu$};
\draw (6.2,0) node {$\,y^\prime_\mu.$};
\draw[<->] (0,-0.4) -- (0,-1.5);
\draw[<->] (4,-0.4) -- (4,-1.5);
\draw (-0.3,0-1) node {$^{T_{S}}$};
\draw (3.7,0-1) node {$^{T_{S}}$};
\draw (-0.2,-2) node {$k^\mu$};
\draw (3.8,-2) node {$\frac{1}{\kappa}\pi_\mu$};
\draw[<->] (0.4,-2) -- (1.9,-0.4);
\draw[<->] (4.4,-2) -- (5.9,-0.4);
\draw (1.4,-1.4) node {$^{T}$};
\draw (5.4,-1.4) node {$^{T}$};
\end{tikzpicture}}\\
\end{center}
For these variables Poisson brackets are involutive on $V_{1}\oplus V^\ast_{2}$ and  for currents  have structural constants carrying the information on the background fluxes
\begin{eqnarray}
&&\{ i_\mu (\sigma), i_\nu (\bar{\sigma}) \} = - 2\kappa B_{\mu \nu \rho} x^{\prime \rho} \delta (\sigma - \bar{\sigma}),\nonumber\\
&&\{ k^\mu (\sigma), k^{\nu} (\bar{\sigma}) \} = -\kappa Q_\rho^{\ \mu \nu} k^\rho \delta(\sigma - \bar{\sigma}) - \kappa^2 R^{\mu \nu \rho} \pi_\rho \delta (\sigma - \bar{\sigma}).
\end{eqnarray}
$H$-flux: $B_{\mu \nu \rho}= \partial_\mu B_{\nu \rho} + \partial_\nu B_{\rho \mu} + \partial_\rho B_{\mu \nu}$ is the Kalb-Ramond field strength,
responsible for non-associativity of a star product in a deformed quantization procedure.
$Q$-flux: $Q_{\rho}^{\ \mu \nu} = \partial_\rho \theta^{\mu \nu}$ determines the noncommutativity parameter dependence on its argument (possibly nonlocal, T-dual, hence nongeometrical).
$R$-flux: $R^{\mu \nu \rho} = \theta^{\mu \sigma} \partial_\sigma \theta^{\nu \rho} + \theta^{\nu \sigma} \partial_\sigma \theta^{\rho \mu} + \theta^{\rho \sigma} \partial_\sigma \theta^{\mu \nu}$ having a form of the generalized Poisson bracket Jacobiator for the open string  noncomutativity parameter.

In this way,
the  initial hamiltonian, expressed in terms of generalized metric
and two symplectic  representations of T-dual theories expressed in terms of currents
are mutually connected
\begin{equation}
{\cal H}=\frac{1}{2\kappa} (X^T)^M H_{MN} X^N=\frac{1}{2 \kappa}  (\hat{X}^T)^M\ G_{MN}\ \hat{X}^N=\frac{1}{2 \kappa}  (\tilde{X}^{T})^M {^\star G}_{MN} \tilde{X}^N,
\end{equation}
 by transformation of generalized phase space coordinates:
\begin{table}[htp]
\centering
\renewcommand\thetable{6} 
\resizebox{\textwidth}{!}{
\begin{tabular}{|l|l|l|}
\hline
\multirow{3}{*}{$2D$ coordinates} & \multirow{3}{*}{$X^M,\;\hat{X}^M,\;\tilde{X}^M$ }&
\multirow{3}{*}{
$\begin{bmatrix}
\kappa x^{\prime \mu} \\
\pi_\mu \\
\end{bmatrix},
\quad
\begin{bmatrix}
\kappa x^{\prime \mu} \\
i_\mu
\end{bmatrix}
=(e^{\hat{B}})^M_{\ N}\ X^N,
\quad
\begin{bmatrix}
k^\mu \\
\pi_\mu 
\end{bmatrix}=(e^{\hat{\theta}})^M_{\ N} X^N $}\\
&&\\
&&\\
\hline
\hline
\multirow{3}{*}{Matrices} & \multirow{3}{*}{$H_{MN},\;G_{MN},\;^\star G_{MN}$} & 
\multirow{3}{*}{$
\begin{bmatrix}
G^E_{\mu \nu} & - 2B_{\mu \rho} (G^{-1})^{\rho \nu} \\
2(G^{-1})^{\mu \rho} B_{\rho \nu} & (G^{-1})^{\mu \nu}
\end{bmatrix},
\quad
\begin{bmatrix}
G_{\mu \nu} & 0 \\
0 & (G^{-1})^{\mu \nu}
\end{bmatrix},
\quad
\begin{bmatrix}
G^E_{\mu \nu} & 0 \\
0 & (G_E^{-1})^{\mu \nu} 
\end{bmatrix}
$}\\
&&\\
&&\\
\hline
\end{tabular}
}
\end{table}\\
while the matrices are related by 
\begin{equation}
( (e^{\hat{B}})^T)_M^{\ K}\ G_{KQ}\ (e^{\hat{B}})^Q_{\ N}=H_{MN},\quad
( (e^{\hat{\theta}})^T )_{M}^{\ L}\ {^\star G}_{LK} (e^{\hat{\theta}})^K_{\ N} =H_{MN}.
\end{equation}
These transformations
diagonalize the generalized metric,
making it dependent on initial and effective metric only.

\section{Geometrical descriptions}
\cleq

In an alternative, a differential geometry description of the string theory
the space is separated into tangent and cotangent parts and one investigates the behaviour of its
constituents for different surroundings. Metric tensor induces the musical isomorphisms
$\flat:TM\rightarrow {T^\ast}\!M,$ $\sharp:{T^\ast}\!M\rightarrow TM,$
or generally
$\flat:E\rightarrow {E^\ast},\quad \sharp:{E^\ast}\rightarrow E$:
\begin{equation}
\eta(X,Y)\equiv\eta(X)(Y)\equiv\eta^\flat(Y),\;\forall X,Y\in E,\quad \eta^\flat:E\rightarrow E^\ast\!,\quad\eta^\flat(Y):E\rightarrow K,
\end{equation}
\begin{equation}
\eta^\sharp\circ\eta^\flat=\hat{1}.
\end{equation}
For different backgrounds and the anchor maps 
governing the projection (inducing the "derivative" ${\cal D}$)
$0\rightarrow T^{\ast}\!M\xrightarrow[\rho^\ast]{} E
\xrightarrow[{\rho}]{\overset{\,j}{\curvearrowleft}}
TM\rightarrow 0$
one finds the involution of splitting is governed by  different brackets.
A structure containing vector bundle $E$, projection to $TM$, scalar product and bracket (multiplication) on $E$ is called algebroid
(bilinear form $\langle X,Y\rangle\in K,\; X\in E^\ast, Y\in E$ naturally induced by a scalar product is also assumed
$\langle(\cdot,Y_{1}),Y_{2}\rangle=(Y_{1},Y_{2})$.
Algebroinds can be doubled to bialgebroids defined on $E\oplus E^\ast$ or specially on $TM\oplus TM^\ast$ with both algebroids carrying their own Lie bracket.
In the enlarged structure \cite{V} one defines 
a bilinear form $(,)$, an extension of a contraction,  such that it annihilates for every two vectors from the same vector space and equals the inner product (contraction) for the mixed terms
\begin{equation}
(x,y)=0,\;\, (\xi,\nu)=0,\;\,(x,\xi)=\langle x,\xi\rangle,
\end{equation}
and naturally extended Lie bracket, based on invariance condition
$(y,\xi)\in K,\;[x, (y,\xi)]=0,\;[\eta, (y,\xi)]=0$,
\begin{equation}
[x,y]=[x,y]_{E},\;\,[\xi,\nu]=[\xi,\nu]_{E^\ast},\;\,[x,\xi]=-ad_{\xi}^\ast x+ad^\ast_{x}\xi,
\end{equation}
where $ad^\ast$ marks the coadjoint related to contraction $\langle,\rangle$ of $ad:x\rightarrow ad_{x}:\, ad_{x}(y) = [x,y]$.
The  pair $(E, E^\ast)$  is sad to be a Lie bialgebroid if $d_{E}$ is a derivation of the Schouten bracket on $E^\ast$.

It all started with Leibniz algebroid, imposing Leibniz rule
\begin{equation}
[\psi,f\psi^\prime]_{E}=f[\psi,\psi^\prime]_{E}+(\rho(\psi)f)\psi^\prime,
\end{equation}
and Leibniz identity 
$[\psi,[\psi^\prime,\psi^{\prime\prime}]_{E}]_{E}=
[[\psi,\psi^\prime]_{E},\psi^{\prime\prime}]_{E}+
[\psi^\prime,[\psi,\psi^{\prime\prime}]_{E}]_{E}$,
which imply the anchor being homeomorphism.
When the bracket is skew-symmetric the Leibniz algebroid becomes the Lie's.
Algebra with  Leibniz identity is called Lodays algebra.

In \cite{Y} the author considers two new conditions connecting the projection $\rho$,
differential $d_{x}$ acting on bilinear form $(\,|\,)$, defined by the bracket
and the bracket itself
\begin{enumerate}\label{eq:pravila}
\item[(i)] $\rho(x)(y|z) = (x | [y, z] + [z, y]) ,$
\item[(ii)] $\rho(x)(y|z) = ([x, y] | z) + (y | [x, z]) ,$
\end{enumerate}
claiming that Leibniz rule and homeomorphism property then follow.
Indeed, simple linearity of $(\,|\,)$ and the rules imply the simple linearity of the bracket and entanglement of existence of Leibniz rule and homeomorphism property,
up to the terms $(\,z|z_{\bot}\,)=0$.

For not skew-symmetric bracket one introduces the mapping ${\cal D}$:
\begin{equation}
[y, z] + [z, y]={\cal D}(y|z),
\end{equation}
and the following left Leibniz rule \cite{JV}
\begin{equation}
[f\psi^\prime,\psi]=f[\psi^\prime,\psi]-(\rho(\psi)f)\psi^\prime+\langle\psi,\psi^\prime\rangle{\cal D}f,
\end{equation}
with ${\cal D}(y|fz):=f{\cal D}(y|z)+{\cal D}f(y|z)$.
The rules for
$y=z$ become
\begin{enumerate}
\item[(i)] $\rho(x)(y|y)=2(x|[y,y])=(x|{\cal D}(y|y)),$
\item[(ii)] $\rho(x)(y|y)=2(y|[x,y]),$
\end{enumerate}
therefore $(x|{\cal D}(y|y))=d_{x}(y|y)$.

In \cite{KS} authors define
a Courant algebroid as a Loday algebroid $(E,\rho, [\cdot,\cdot])$ together with an
invariant $E$-metric ${^{\scriptscriptstyle E}}\!g$, where invariance implies
\begin{equation}
{^{\scriptscriptstyle E}}\!g([\psi,\psi],\psi^\prime)=\frac{1}{2}
\rho(\psi^\prime){^{\scriptscriptstyle E}}\!g(\psi,\psi).
\end{equation}

While Lie and Koszul bracket are relevant for 
tangent
and cotangent spaces, the second being induced by Poisson structure,
for the sum of tangent and cotangent spaces, named generalized geometry \cite{NH},
Courant introduced the bracket
\begin{equation}
 [X_1 +\xi_1,X_2 +\xi_2] = [X_1,X_2]+L_{X_1}\xi_2-L_{X_2}\xi_1+
d\Big(\frac{1}{2}\big(\xi_1(X_2)-\xi_2(X_1)\big)\Big)
\end{equation}
for which the  Leibniz identity  is not satisfied in general.
The discrepancy from Leibniz identity named  the Leibniz anomaly $[e_1,fe_2]-f[e_1,e_2]-(\rho(e_1)f)e_2$
vanishes 
for Dirac structures.
The connection between spaces is however given by 
contraction, natural parring, generalized for bialgebroid.
The definitions and aspects of Courant's paper \cite{Kurantov rad} were discussed  in \cite{MT}.

This bracket is used to define a Courant algebroid, but the definition is not unique,
for there is a definition trough Dorfman bracket, which is not antisymmetric.
Both definitions of Courant algebroid are disscussed in \cite{RoytenbergPHD}.
In \cite{PBV} one recols the definition of Courant algebroid with Dorfmann bracket, given by the following five rules:
\begin{enumerate}
\item Leibniz property or left Jacoby identity $[e_1, [e_2, e_3]] = [[e_1,e_2], e_3] + [e_2, [e_1, e_2]]$
\item Anchor being a homomorphism $\rho [e_1, e_2] = [\rho(e_1), \rho(e_2)]$
\item Leibniz rule $[e_1, fe_2] = f[e_1, e_2] + {\cal L}_{\rho(e_1)}(f)e_2$
\item $[e, e] = \frac{1}{2}{\cal D}\langle e, e\rangle$, with ${\cal D}=\rho^\ast d$
\item Self-adjointness (following from the concept of an invariant form) $\rho(e_1)\langle e_2, e_3\rangle = \langle[e_1, e_2], e_3\rangle + \langle e_2, [e_1, e_3]\rangle$.
\end{enumerate}

The definition trough Courant bracket is also composed of five rules.
The Courant algebroid is a vector bundle, with an antisymmetric bracket, bilinear form and projection to the tangent space,
satisfying the following:
Jacobiator for the bracket equals ${\cal D}T$ where $T(e_1,e_2,e_3) =\frac{1}{3}([e_1,e_2],e_3) + cyc$
and ${\cal D}=\frac{1}{2}\beta^{-1}\rho^\ast d_{0}$.
The projection is a homeomorphism.
The  Leibniz anomaly is $-(e_{1},e_{2}){\cal D}f$ and
\begin{eqnarray}
&&\rho\circ{\cal D}=0,\quad\text{ie.}\quad \big({\cal D}f,{\cal D}g\big)=0,\nonumber\\
&&\rho(e)(h_1,h_2) = ([e,h_1] + {\cal D}(e,h_1),h_2) + (h_1,[e,h_2] + {\cal D}(e,h_2)).
\end{eqnarray}
Both Leibniz identity and rule do not stand for antisymmetric bracket \cite{KS},
and by \cite{Alekseev,MT} the transition to antisymmetric bracket,
transforms the Leibniz identity  into the first rule.


\subsection{Physical manifestation}

The same objects, in a different context and not necessarily with the exact name are present in physical literature.

In the paper \cite{BDPR} authors consider the quasi Poisson structure
\begin{equation}
\{f,g\}=\beta^{ij}\partial_{i}f\,\partial_{j}g.
\end{equation}
It is assumed that it gives rise to natural mapping, allowing 
existence  of derivatives 
$\beta^{ij}\partial_j$ acting
in a dual theory, along with derivatives  over the dual coordinates $\tilde\partial^{i}$.
Commutators of these two types of derivatives
reveal background fluxes, and the implosion of Jacobi identity for all combinations
gives Bianchi identities.
Geometric flux is generated by the consideration of 
the fielbeins as vector field parameters and their duals.

In \cite{HHZmulti},
the double string field theory is considered,
formulated in constant background with fluctuations and given 
in terms of derivatives combining both coordinates of doubled configuration space.
In such description,
generalized doubled space-time coordinate transformations include both diffeomorphisms and B-field gauge transformation.
Contrary to ordinary diffeomorphisms, the composition of generalized displays nonassociativity.
Additionally the gauge transformations of background fields
can not be presented in terms of Lie derivatives on the doubled space 
but rather by generalized Lie derivatives defined by $\delta_\xi{\cal H}_{MN}=\hat{\cal L}_\xi{\cal H}_{MN}$.
Consequently,
the Lie derivatives are 
replaced by the generalized Lie derivatives or Dorfman bracket and the Courant bracket by the
C-bracket.

C--bracket was introduced \cite{SIEG},
  for commutation relations of group elements $e^{-i\Lambda}$ representing general coordinate transformations given in terms of basis of oscillators for heterotic string
\begin{equation}
 \lambda^{M}_{[1,2]}=\lambda^{N}_{[1}\partial_{N}\lambda^{M}_{2]}
-\frac{1}{2}\lambda^{N}_{[1}\partial^{M}\lambda_{2]N},
\end{equation} and was named new Lie derivative.

Poisson bracket's of generalized currents, formed for the effective variables of two representations of initial theory,
was show in \cite{PrviRad} to reproduce the Courant bracket
 twisted by Kalb-Ramond field $B_{\mu\nu}$ and the Roytenberg bracket,
demonstrating their T-duality, for  Roytenberg bracket
can be seen as Courant's bracket twisted by a bi-vector $\kappa\theta^{\mu\nu}$.
Roytenberg introduced the bracket in \cite{Roytenberg} afterwards named Roytenberg's,
as a version of a Koszul bracket (\ref{eq:kozbra}),
by
\begin{equation}
[X.Y]_{\gamma,\omega}=
{\cal L}^{\gamma}_{\tilde\omega X}Y
-{\cal L}^{\gamma}_{\tilde\omega Y}X
-d_\gamma\big(\omega(X,Y)\big),
\end{equation}
 later named twisted Courant.
In the same paper author discusses the linearisation of derivatives,
defined by brackets and the twists implied by the Hamiltonian fields on a Loday's bialgebra.
For structures to remain within same categories one imposes versions of 
Mauer-Cartan equation. This is analogous to conditions for restriction
to Dirac structures.

The rotations in the 2D generalized phase space, $B$- and $\theta$-transforms, within symplectic T-duality,
make change of  the symmetry generators
\begin{equation}
{\cal G}_{GCT} (\xi) + {\cal G}_{LG} (\lambda) =
\int_0^{2\pi} d\sigma\Big[\xi^\mu\pi_\mu+ \lambda_\mu \kappa x^{\prime\mu} \Big] = \int_0^{2\pi} d\sigma (\Lambda^T)^M \Omega_{MN} X^N.
\end{equation}
In \cite{DrugiRad} the Lie bracket  appearing within the algebra
of general coordinate transformation generator was shown to generalizes into  a Courant bracket within self T-dual symmetry generator's algebra.
Inclusion of the Kalb-Ramond field, changes the algebra of the generators
so that their parameters form the B-twisted Courant bracket,  which in self T-dual rearrangement  becomes a $\theta$-twisted Courant bracket.
In \cite{TreciRad}, the  twist of Courant bracket  by both B and $\theta$ simultaneously was performed. In
double theory, the algebra of the analogues of the above generators was found, containing C-bracket
together with B and $\theta$ twisted C-brackets \cite{CetvrtiRad}. The algebroidal properties of these brackets 
are discussed in \cite{PHDIvanisevic}.


\section{The C-bracket generator}
\cleq

Let us consider symmetry generators of double bosonic string  theory, consisting of two mutualy T-dual thoeries,
defined on doubled phase space, 
parametrically dependent on doubled coordinate $(x,y)$,
partially projected by T-duality
to the single phase space $(x,\pi)$.

The generators
\begin{equation}\label{eq:gener}
G=\int_{0}^{2\pi}d\sigma\,{\cal G},\;{\cal G}=\begin{bmatrix} \xi^\mu(x,y) & \lambda_\mu(x,y)\end{bmatrix} \begin{bmatrix} \pi_\mu\\  {^\star \pi_\mu}\end{bmatrix},
\end{equation}
are analogue of single
generators $G=\int_{0}^{2\pi}d\sigma\,{\cal G}=\int_{0}^{2\pi}d\sigma\,\lambda^\mu(x)\pi_\mu$.
By a T-duality projection  to the phase space of the initial theory
(keeping in mind both theories or both sets of background fields) one obtains
\begin{equation}
{\cal G}=\begin{bmatrix} \xi^\mu(x,\frac{1}{\kappa}P) & \lambda_\mu(x,\frac{1}{\kappa}P)\end{bmatrix} \begin{bmatrix} \pi_\mu\\  \kappa x^{\prime\mu}\end{bmatrix}
=\Lambda^{T}\Pi,
\end{equation}
with $$P_\mu(\sigma)=\int d\sigma\,\pi_\mu(\sigma)\equiv\int^{\sigma}d\eta\,\pi_\mu(\eta),$$
representing the double coordinate.
Their Poisson bracket $\{\cdot,\ast\}=\int d\eta\Big{\{}\frac{\partial\,\cdot}{\partial x^\varepsilon}\frac{\partial\ast}{\partial \pi_\varepsilon}-
\frac{\partial\,\cdot}{\partial \pi_\varepsilon}\frac{\partial\ast}{\partial x^\varepsilon}\Big{\}}$,
is calculated 
part by part, assuming the standard Poisson brackets for the phase space variables.
We use the fact $\delta$--function is symmetric,
and the step function $\theta(\sigma)=\int d\sigma\delta(\sigma)$
obeys
the relation $\int_{0}^{\sigma} d\eta f(\eta)\delta(\eta-\bar\sigma)=f(\bar\sigma)\big(\theta(\sigma-\bar{\sigma})+\theta(\bar{\sigma})\big)$.

The generators Poisson bracket reads
\begin{eqnarray}\label{eq:PBG}
\big{\{}{\cal G}(\sigma),{\cal G}(\bar\sigma)\big{\}}&=&
\big{(}
(\xi^\nu_{2}\partial_\nu\xi^\mu_{1}-\xi^\nu_{1}\partial_\nu\xi^\mu_{2}
+\lambda_{2\nu}\partial^\nu\xi^\mu_{1}-\lambda_{1\nu}\partial^\nu\xi^\mu_{2})\pi_\mu
\nonumber\\
&+&
(
\xi^\nu_{2}\partial_\nu\lambda_{1\mu}-\xi^\nu_{1}\partial_\nu\lambda_{2\mu}
+\lambda_{2\nu}\partial^\nu\lambda_{1\mu}-\lambda_{1\nu}\partial^\nu\lambda_{2\mu}
)\kappa x^{\prime\mu}
\big{)}\delta(\sigma-\bar\sigma)
\nonumber\\
&+&\big{(}
\xi_{1}^\mu(\sigma)\lambda_{2\mu}(\bar\sigma)+
\xi_{2}^\mu(\bar\sigma)\lambda_{1\mu}(\sigma)
\big{)}\delta^\prime(\sigma-\bar\sigma)
\nonumber\\
&+&\frac{1}{\kappa}\partial_\rho\xi^{\mu}_{1}\partial^\rho\xi_{2}^{\nu}\pi_\mu(\sigma)\pi_\nu(\bar\sigma)
-\frac{1}{\kappa}\partial^\rho\xi^{\mu}_{1}\partial_\rho\xi_{2}^{\nu}\pi_\mu(\sigma)\pi_\nu(\bar\sigma)
\nonumber\\
&+&\partial_\rho\xi^\mu_{1}\partial^\rho\lambda_{2\nu}\pi_\mu(\sigma)x^{\prime\nu}(\bar\sigma)
-\partial^\rho\xi^\mu_{1}\partial_\rho\lambda_{2\nu}\pi_\mu(\sigma)x^{\prime\nu}(\bar\sigma)
\nonumber\\
&-&\partial_\rho\xi^\mu_{2}\partial^\rho\lambda_{1\nu}\pi_\mu(\bar\sigma)x^{\prime\nu}(\sigma)
+\partial^\rho\xi^\mu_{2}\partial_\rho\lambda_{1\nu}\pi_\mu(\bar\sigma)x^{\prime\nu}(\sigma)
\nonumber\\
&+&\kappa\partial_\rho\lambda_{1\mu}\partial^\rho\lambda_{2\nu}x^{\prime\mu}(\sigma)x^{\prime\nu}(\bar\sigma)
-\kappa\partial^\rho\lambda_{1\mu}\partial_\rho\lambda_{2\nu}x^{\prime\mu}(\sigma) x^{\prime\nu}(\bar\sigma).
\end{eqnarray}

By \cite{Alekseev} current algebra induces the definition of a bracket (multiplication)
of generator parameters and their bilinear scalar product in the anomalous term 
($\sim\delta^\prime$). Analogously, here one defines
\begin{equation}\label{eq:SPA}
\big(\Lambda_{1}(\sigma),\Lambda_{2}(\bar\sigma)\big)=\xi_{1\mu}(\sigma)\lambda_{2}^\mu(\bar\sigma)+
\lambda_{1}^\mu(\sigma)\xi_{2\mu}(\bar\sigma).
\end{equation}
The structure reduces to Dirac structure in absence of the anomaly.
The absence could be achieved if parameters $\xi$ and $\lambda$
 have non-zero values for  disjunct choices of directions $\mu$.

Rewriting the equation  (\ref{eq:PBG}) in the matrix notation,
one obtains the term proportional to delta function $\delta(\sigma-\bar\sigma)$ in a form 
\begin{equation}
\big[\Lambda^{T}_{2}\big(\boldsymbol{\partial} \Lambda^{T}_{1}\big)-\Lambda^{T}_{1}\big(\boldsymbol{\partial} \Lambda^{T}_{2}\big)\big]\Pi,
\end{equation}
antisymmetric in $1\leftrightarrow 2$,
with $\boldsymbol{\partial}=\begin{bmatrix} \partial_\mu\\  \partial^{\mu}\end{bmatrix}$.
It defines a 
bracket defining how generator parameters fit into the resulting parameter,
within a linear in $\Pi$ term
\begin{equation}\label{eq:2bra}
\big[\Lambda_{1},\Lambda_{2}\big]
=\big(\Lambda_{2},\boldsymbol{\partial}\big)\Lambda_{1}
-\big(\Lambda_{1},\boldsymbol{\partial}\big)
\Lambda_{2},
\end{equation}
where 
\begin{equation}
\big(\Lambda,\boldsymbol{\partial}\big)=\xi^\varepsilon\,\partial_{\varepsilon}+\lambda_{\varepsilon}\,\partial^\varepsilon\,.
\end{equation}

The term proportional to the delta function $\sigma$-derivative $\delta^\prime(\sigma-\bar\sigma)$
can be partially integrated, using $\delta^\prime(\sigma-\bar\sigma)=\frac{1}{2}\delta^\prime(\sigma-\bar\sigma)-\frac{1}{2}\delta^\prime(\bar\sigma-\sigma)$,
so that in the matrix notation it becomes
\begin{equation}
\frac{1}{2}\Lambda^{T}_{1}(\Pi^{T}\boldsymbol{\partial})\Omega\Lambda_{2}
-
\frac{1}{2}\Lambda^{T}_{2}(\Pi^{T}\boldsymbol{\partial})\Omega\Lambda_{1},
\end{equation}
also antisymmetric in $1\leftrightarrow 2$, now being the term proportional to 
$\delta(\sigma-\bar\sigma)$, therefore making change of the bracket.
What is left in equation (\ref{eq:PBG}) equals
\begin{equation}
\frac{1}{\kappa}{\scriptstyle{\boldsymbol{\partial}}^{T}}\Lambda^{T}_{1}
\Pi\;{\scriptstyle \bar\Omega\,\boldsymbol{\partial}}\,\Lambda^{T}_{2}\Pi,
\end{equation}
with $\Omega=\begin{bmatrix}0&I\\
I&0\end{bmatrix},\;\bar\Omega=\begin{bmatrix}0&I\\
-I&0\end{bmatrix}$.

If the parameters in the generator (\ref{eq:gener}) depend on initial coordinate only, then their Poisson algebra in a initial phase space
does not contain terms originating from $P_\mu$, which except for the $\theta$-dependent terms contribute
by terms proportional to
\begin{equation}
\{P_\mu,x^{\prime\nu}\}=-\delta_\mu^\nu\delta(\sigma-\bar\sigma).
\end{equation}
Then the algebra is just
\begin{eqnarray}
\big{\{}{\cal G}(\sigma),{\cal G}(\bar\sigma)\big{\}}&=&
\big{(}
(\xi^\nu_{2}\partial_\nu\xi^\mu_{1}-\xi^\nu_{1}\partial_\nu\xi^\mu_{2}
)\pi_\mu
+(
\xi^\nu_{2}\partial_\nu\lambda_{1\mu}-\xi^\nu_{1}\partial_\nu\lambda_{2\mu}
)\kappa x^{\prime\mu}
\big{)}\delta(\sigma-\bar\sigma)
\nonumber\\
&+&\big{(}
\xi_{1}^\mu(\sigma)\lambda_{2\mu}(\bar\sigma)+
\xi_{2}^\mu(\bar\sigma)\lambda_{1\mu}(\sigma)
\big{)}\delta^\prime(\sigma-\bar\sigma).
\end{eqnarray}
The form of the scalar product (\ref{eq:SPA}) is unchanged,
but there is only $x$-dependence. 
If the anomaly term is partially integrated one obtains the change of the bracket by
\begin{equation}
\frac{1}{2}\Lambda^{T}_{1}(x^{\prime}\partial)\Omega\Lambda_{2}
-
\frac{1}{2}\Lambda^{T}_{2}(x^{\prime}\partial)\Omega\Lambda_{1}.
\end{equation}

However, one need not do the symmetrized partial integration but instead
bring the anomaly term to the form
\begin{equation}
\Big(\Lambda_{1}(\sigma),\Lambda_{2}(\sigma)\Big)
\delta^\prime(\sigma-\bar\sigma)=
\big(\xi_{1\mu}(\sigma)\lambda_{2}^\mu(\sigma)+
\lambda_{1}^\mu(\sigma)\xi_{2\mu}(\sigma)
\big)\delta^\prime(\sigma-\bar\sigma).
\end{equation}
Now, the anomaly can be cancelled by the appropriate functional dependence between the  parameters \cite{IS}.
The resulting generator would change by
$\Lambda^{T}_{1}(\Pi^{T}\boldsymbol{\partial})\Omega\Lambda_{2}$ or
$\Lambda^{T}_{1}(x^\prime{\partial})\Omega\Lambda_{2}$
and the bracket would no longer be antisymmetric.

To conclude, either the bracket is Lie's and the anomaly is two parameter function,
or the bracket is C-bracket, antisymmetric or $D$ \cite{Dbrack}, and the anomaly is one parameter function.


\section{3-bracket anomaly}
\cleq
Poisson bracket with yet another generator $\Big{\{}\big{\{}{\cal G}(\sigma),{\cal G}(\bar\sigma)\big{\}},{\cal G}(\bar\eta)\Big{\}}$
is more easily calculated in a matrix notation, with the use of
\begin{eqnarray}
&&\partial_\varepsilon\Pi(\sigma)=\begin{bmatrix}0\\\kappa \delta^{\prime}(\sigma-\eta)\delta^{\mu}_\varepsilon\end{bmatrix},
\qquad
\partial^\varepsilon\Pi(\sigma)=\begin{bmatrix}\delta(\eta-\sigma)\delta_{\mu}^\varepsilon\\0\end{bmatrix},
\nonumber\\
&&\partial^\varepsilon(\eta)\Lambda(\sigma)=\frac{1}{\kappa}\partial^\varepsilon\Lambda(\sigma)\theta(\sigma-\eta).
\end{eqnarray}

It defines the 3-bracket within linear in  $\Pi\delta^2(\sigma,\bar\sigma,\bar\eta)$  term
\begin{eqnarray}
\big[\big[\Lambda_{1},\Lambda_{2}\big],\Lambda_{3}\big]&=&
\big(\Lambda_{3},\boldsymbol{\partial}\big)\big(\Lambda_{2},\boldsymbol{\partial}\big)\Lambda_{1}
-
\big(\Lambda_{3},\boldsymbol{\partial}\big)\big(\Lambda_{1},\boldsymbol{\partial}\big)\Lambda_{2}
\nonumber\\
&+&\big(\Lambda_{3},\big(\Lambda_{2},\boldsymbol{\partial}\big)\boldsymbol{\partial}\big{)}\Lambda_{1}
-
\big(\Lambda_{3},\big(\Lambda_{1},\boldsymbol{\partial}\big)\boldsymbol{\partial}\big)\Lambda_{2}
\nonumber\\
&-&\big(\Lambda_{2},\boldsymbol{\partial}\big)\big(\Lambda_{1},\boldsymbol{\partial}\big)\,\Lambda_{3}
+\big(\Lambda_{1},\boldsymbol{\partial}\big)\big(\Lambda_{2},\boldsymbol{\partial}\big)\,\Lambda_{3}
\end{eqnarray}
which is  connected to the 2-bracket (\ref{eq:2bra}) by
\begin{eqnarray}\label{eq:23bra}
\big[\big[\Lambda_{1},\Lambda_{2}\big],\Lambda_{3}\big]&=&
\big(\Lambda_{3},\boldsymbol{\partial}\big)\big[\Lambda_{1},\Lambda_{2}\big]
-\big(\big[\Lambda_{1},\Lambda_{2}\big],\boldsymbol{\partial}\big)\,\Lambda_{3},
\end{eqnarray}
confirming that the 3-bracket is in accordance with the 2-bracket (\ref{eq:2bra}).
Jacoby identity for this 3-bracket is satisfied.
Ie. anomaly free part is Lie's.
The relation (\ref{eq:23bra}) will differ for different choices of $2$-bracket, but the Jacobi 
identity stands.

The anomalous term is now more demanding and it consists of terms linear in $\Pi$
and purely $\Lambda$ dependent terms:
\begin{eqnarray}
&&\delta^\prime(\sigma-\bar\eta)\theta(\bar\sigma-\sigma)\;\Big{[}
\big(\Lambda_{3},{\partial}_\beta\Lambda_{1}\big){\partial}^\beta\Lambda_{2}^{T}
+\big(\Lambda_{3},{\partial}^\beta\Lambda_{1}\big){\partial}_\beta\Lambda_{2}^{T}\Big{]}\Pi
\nonumber\\
&+&
\delta^\prime(\bar\sigma-\bar\eta)\theta(\bar\sigma-\sigma)\;\Big{[}
\big(\Lambda_{3},{\partial}_\beta\Lambda_{2}\big){\partial}^\beta\Lambda_{1}^{T}
+\big(\Lambda_{3},{\partial}^\beta\Lambda_{2}\big){\partial}_\beta\Lambda_{1}^{T}\Big{]}\Pi
\nonumber\\
&-&\delta^\prime(\sigma-\bar\sigma)\theta(\sigma-\bar\eta)\;
\big(\Lambda_{2},\partial^\beta\Lambda_{1}\big)\partial_\beta\Lambda_{3}^{T}\Pi
-\delta^\prime(\sigma-\bar\sigma)\theta(\bar\sigma-\bar\eta)\;
\big(\Lambda_{1},\partial^\beta\Lambda_{2}\big)\partial_\beta\Lambda_{3}^{T}\Pi
\nonumber\\
&+&\delta^\prime(\sigma-\bar\sigma)\delta(\sigma-\bar\eta)\;
\Big[\kappa\big(\Lambda_{2},\big(\Lambda_{3},\boldsymbol{\partial}\big)\Lambda_{1}\big)
-\big(\Lambda_{2},\partial_\beta\Lambda_{1}\big)\partial^\beta\Lambda_{3}^{T}\Pi\Big]
\nonumber\\
&+&
\delta^\prime(\sigma-\bar\sigma)\delta(\bar\sigma-\bar\eta)\;
\Big[\kappa\big(\Lambda_{1},\big(\Lambda_{3},\boldsymbol{\partial}\big)\Lambda_{2}\big)
-\big(\Lambda_{1},\partial_\beta\Lambda_{2}\big)\partial^\beta\Lambda_{3}^{T}\Pi\Big]
\nonumber\\
&+&\delta^\prime(\sigma-\bar\eta)\delta(\sigma-\bar\sigma)\;
\kappa\big(\Lambda_{3},\big(\Lambda_{2},\boldsymbol{\partial}\big)\Lambda_{1}\big)
-\delta^\prime(\bar\sigma-\bar\eta)\delta(\sigma-\bar\sigma)\;
\kappa\big(\Lambda_{3},\big(\Lambda_{1},\boldsymbol{\partial}\big)\Lambda_{2}\big).\nonumber\\
\end{eqnarray}

Because $\theta$ dependent terms contain two derivatives each, assuming linear dependence on arguments a partial integration, here not admissible to symmetrization,
produces  $\delta^2(\sigma,\bar\sigma,\bar\eta)$  dependent terms
\begin{eqnarray}
 &&\delta^2(\sigma,\bar\sigma,\bar\eta)\;\Big{\{}
\big(\Lambda_{3},{\partial}_\beta\Lambda_{1}\big){\partial}^\beta\Lambda_{2}^{T}
+\big(\Lambda_{3},{\partial}^\beta\Lambda_{1}\big){\partial}_\beta\Lambda_{2}^{T}
\nonumber\\
&&\qquad
-\big(\Lambda_{3},{\partial}_\beta\Lambda_{2}\big){\partial}^\beta\Lambda_{1}^{T}
-\big(\Lambda_{3},{\partial}^\beta\Lambda_{2}\big){\partial}_\beta\Lambda_{1}^{T}
\nonumber\\
&&\qquad+\Big[
\big(\Lambda_{2},\partial^\beta\Lambda_{1}\big)
-\big(\Lambda_{1},\partial^\beta\Lambda_{2}\big)
\Big]\partial_\beta\Lambda_{3}^{T}
\Big{\}}\Pi.
\end{eqnarray}
Jacobiator for this addition would be
$$\partial_{\mu-}\big(\Lambda_{1},\Lambda_{3}\big)\partial^\mu\Lambda_{2}
+\partial_{\mu-}\big(\Lambda_{3},\Lambda_{2}\big)\partial^\mu\Lambda_{1}
+\partial_{\mu-}\big(\Lambda_{2},\Lambda_{1}\big)\partial^\mu\Lambda_{3},$$
with $\partial_{\mu-}\big(\Lambda_{1},\Lambda_{3}\big)=
\big(\partial_{\mu}\Lambda_{1},\Lambda_{3}\big)-
\big(\Lambda_{1},\partial_{\mu}\Lambda_{3}\big)$.

All other anomaly terms,
containing just one derivative will contribute after the action of the 
$\sigma,\bar\sigma$ or $\bar\eta$ derivative, forming 
 \begin{equation}
{\cal D}=\frac{1}{\kappa}\big(\Pi,\boldsymbol{\partial}\big),
\end{equation}
dependent on the world-sheet parameters it acts on.
Depending on the representation
$\delta^\prime(\sigma-\bar\sigma)=-\delta^\prime(\bar\sigma-\sigma)=
\frac{1}{2}\delta^\prime(\sigma-\bar\sigma)-\frac{1}{2}\delta^\prime(\bar\sigma-\sigma)$
 one out of three expressions is obtained
\begin{eqnarray}
&&-\kappa\big(\Lambda_{2},{\cal D}\big(\Lambda_{3},\boldsymbol{\partial}\big)\Lambda_{1}\big)
\quad\text{or}\quad
\kappa\big({\cal D}\Lambda_{2},\big(\Lambda_{3},\boldsymbol{\partial}\big)\Lambda_{1}\big)
\nonumber\\
&&\quad\text{or}\quad
-\frac{\kappa}{2}\big(\Lambda_{2},{\cal D}\big(\Lambda_{3},\boldsymbol{\partial}\big)\Lambda_{1}\big)
+\frac{\kappa}{2}\big({\cal D}\Lambda_{2},\big(\Lambda_{3},\boldsymbol{\partial}\big)\Lambda_{1}\big),
\end{eqnarray}
\begin{eqnarray}
&&\kappa\big(\Lambda_{1},{\cal D}\big(\Lambda_{3},\boldsymbol{\partial}\big)\Lambda_{2}\big)
\quad\text{or}\quad
-\kappa\big({\cal D}\Lambda_{1},\big(\Lambda_{3},\boldsymbol{\partial}\big)\Lambda_{2}\big)
\nonumber\\
&&\quad\text{or}\quad
\frac{\kappa}{2}\big(\Lambda_{1},{\cal D}\big(\Lambda_{3},\boldsymbol{\partial}\big)\Lambda_{2}\big)
-\frac{\kappa}{2}\big({\cal D}\Lambda_{1},\big(\Lambda_{3},\boldsymbol{\partial}\big)\Lambda_{2}\big),
\end{eqnarray}
\begin{eqnarray}
&&-\kappa\big(\Lambda_{3},{\cal D}\big[\Lambda_{1},\Lambda_{2}\big]\big)
\quad\text{or}\quad
\kappa\big({\cal D}\Lambda_{3},\big[\Lambda_{1},\Lambda_{2}\big]\big)
\nonumber\\
&&\quad\text{or}\quad
-\frac{\kappa}{2}\big(\Lambda_{3},{\cal D}\big[\Lambda_{1},\Lambda_{2}\big]\big)
+\frac{\kappa}{2}\big({\cal D}\Lambda_{3},\big[\Lambda_{1},\Lambda_{2}\big]\big).
\end{eqnarray}
Total derivatives are assumed to be cancelled for the closed string.
In this way the anomaly is cancelled.
If one considers the generator density only,
then the same expressions changing the bracket are obtained using
\begin{equation}
f(\bar\sigma)\delta^\prime(\sigma-\bar\sigma)=f^\prime(\sigma)\delta(\sigma-\bar\sigma)+
f(\sigma)\delta^\prime(\sigma-\bar\sigma)
\end{equation}
and the anomaly is still present but it becomes function of one variable.

Uniting all representations, the Jacobiator equals
\begin{eqnarray}
&&\kappa{\cal D}\big(
\Lambda_{1},\beta\big(\Lambda_{3},\boldsymbol{\partial}\big)\Lambda_{2}+\gamma\big(\Lambda_{2},\boldsymbol{\partial}\big)\Lambda_{3}
\big)
\nonumber\\
&&
+\,\kappa{\cal D}\big(
\Lambda_{2},\beta\big(\Lambda_{1},\boldsymbol{\partial}\big)\Lambda_{3}+\gamma\big(\Lambda_{3},\boldsymbol{\partial}\big)\Lambda_{1}
\big)
\nonumber\\
&&+\,\kappa{\cal D}\big(
\Lambda_{3},\beta\big(\Lambda_{2},\boldsymbol{\partial}\big)\Lambda_{1}+\gamma\big(\Lambda_{1},\boldsymbol{\partial}\big)\Lambda_{2}
\big)
\end{eqnarray}
with
$\beta=b-c,\;\gamma=c-a$ where $a,b,c\in\big{\{}0,\frac{1}{2},1\big{\}}$.
Generalizing the bracket
\begin{equation}
\big[\Lambda_{1},\Lambda_{2}\big]
=\big(\Lambda_{2},\boldsymbol{\partial}\big)\Lambda_{1}
-\big(\Lambda_{1},\boldsymbol{\partial}\big)
\Lambda_{2},
\end{equation}
into a linear combination of its factors 
\begin{equation}\label{eq:gbr}
\big[\Lambda_{1},\Lambda_{2}\big]_{(\beta,\gamma)}
=\beta\big(\Lambda_{2},\boldsymbol{\partial}\big)\Lambda_{1}
+\gamma\big(\Lambda_{1},\boldsymbol{\partial}\big)
\Lambda_{2},
\end{equation}
the Jacobiator is rewritten as
\begin{eqnarray}
&&\kappa{\cal D}\Big{[}\Big(
\Lambda_{1},\big[\Lambda_{2},\Lambda_{3}\big]_{(\beta,\gamma)}
\Big)
+\,\Big(
\Lambda_{2},\big[\Lambda_{3},\Lambda_{1}\big]_{(\beta,\gamma)}
\Big)
+\,\Big(
\Lambda_{3},\big[\Lambda_{1},\Lambda_{2}\big]_{(\beta,\gamma)}
\Big)\Big{]}.
\end{eqnarray}

If the generator parameters are restricted to depend on the initial coordinates only, the anomaly becomes
\begin{eqnarray}
&-&\delta^\prime(\bar\sigma-\sigma)\delta(\sigma-\bar\eta)\;
\kappa\big(\Lambda_{2},
\xi_{3}^\mu\partial_\mu
\Lambda_{1}\big)
\nonumber\\
&+&
\delta^\prime(\sigma-\bar\sigma)\delta(\bar\sigma-\bar\eta)\;
\kappa\big(\Lambda_{1},
\xi_{3}^\mu\partial_\mu
\Lambda_{2}\big)
\nonumber\\
&-&\delta^\prime(\bar\eta-\sigma)\delta(\sigma-\bar\sigma)\;
\kappa\big(\Lambda_{3},
\xi_{2}^\mu\partial_\mu
\Lambda_{1}-\xi_{1}^\mu\partial_\mu
\Lambda_{2}\big).
\end{eqnarray}
Partially integrating one obtains
\begin{eqnarray}
&&-\kappa\big(\Lambda_{2},{\cal D}\big(\xi_{3},{\partial}\big)\Lambda_{1}\big)
\quad\text{or}\quad
\kappa\big({\cal D}\Lambda_{2},\big(\xi_{3},{\partial}\big)\Lambda_{1}\big)
\nonumber\\
&&\quad\text{or}\quad
-\frac{\kappa}{2}\big(\Lambda_{2},{\cal D}\big(\xi_{3},{\partial}\big)\Lambda_{1}\big)
+\frac{\kappa}{2}\big({\cal D}\Lambda_{2},\big(\xi_{3},{\partial}\big)\Lambda_{1}\big),
\end{eqnarray}
\begin{eqnarray}
&&\kappa\big(\Lambda_{1},{\cal D}\big(\xi_{3},{\partial}\big)\Lambda_{2}\big)
\quad\text{or}\quad
-\kappa\big({\cal D}\Lambda_{1},\big(\xi_{3},{\partial}\big)\Lambda_{2}\big)
\nonumber\\
&&\quad\text{or}\quad
\frac{\kappa}{2}\big(\Lambda_{1},{\cal D}\big(\xi_{3},{\partial}\big)\Lambda_{2}\big)
-\frac{\kappa}{2}\big({\cal D}\Lambda_{1},\big(\xi_{3},{\partial}\big)\Lambda_{2}\big),
\end{eqnarray}
\begin{eqnarray}
&&-\kappa\big(\Lambda_{3},{\cal D}\big[\Lambda_{1},\Lambda_{2}\big]\big)
\quad\text{or}\quad
\kappa\big({\cal D}\Lambda_{3},\big[\Lambda_{1},\Lambda_{2}\big]\big)
\nonumber\\
&&\quad\text{or}\quad
-\frac{\kappa}{2}\big(\Lambda_{3},{\cal D}\big[\Lambda_{1},\Lambda_{2}\big]\big)
+\frac{\kappa}{2}\big({\cal D}\Lambda_{3},\big[\Lambda_{1},\Lambda_{2}\big]\big),
\end{eqnarray}
where
\begin{equation}
\big[\Lambda_{1},\Lambda_{2}\big]
=\big(\xi_{2},{\partial}\big)\Lambda_{1}
-\big(\xi_{1},{\partial}\big)
\Lambda_{2},
\end{equation}
with $\big(\xi,{\partial}\big)=\xi^\varepsilon\,\partial_{\varepsilon}\,$
and ${\cal D}$ reduced to
${\cal D}\,\ast=\frac{1}{\kappa}\big(\kappa x^\prime,{\partial}\big)\,\ast=x^{\prime\varepsilon}
\partial_{\varepsilon}\,\ast
=(\ast)^\prime$.

However, the contribution to the Jacobiator in
$x$-dependent case is just a total derivative.
It equals
\begin{eqnarray}
&&\kappa{\cal D}\big(
\Lambda_{1},\beta\big(\xi_{3},\partial\big)\Lambda_{2}+\gamma\big(\xi_{2},\partial\big)\Lambda_{3}
\big)
\nonumber\\
&&
+\,\kappa{\cal D}\big(
\Lambda_{2},\beta\big(\xi_{1},\partial\big)\Lambda_{3}+\gamma\big(\xi_{3},\partial\big)\Lambda_{1}
\big)
\nonumber\\
&&+\,\kappa{\cal D}\big(
\Lambda_{3},\beta\big(\xi_{2},\partial\big)\Lambda_{1}+\gamma\big(\xi_{1},\partial\big)\Lambda_{2}
\big)
\end{eqnarray}
and defining  the generalized bracket, reduction of (\ref{eq:gbr}) by
\begin{equation}
\big[\Lambda_{1},\Lambda_{2}\big]_{(\beta,\gamma)}
=\beta\big(\xi_{2},{\partial}\big)\Lambda_{1}
+\gamma\big(\xi_{1},{\partial}\big)
\Lambda_{2},
\end{equation}
it is rewritten as 
$
\kappa\Big{[}\Big(
\Lambda_{1},\big[\Lambda_{2},\Lambda_{3}\big]_{(\beta,\gamma)}
\Big)
+\,\Big(
\Lambda_{2},\big[\Lambda_{3},\Lambda_{1}\big]_{(\beta,\gamma)}
\Big)
+\,\Big(
\Lambda_{3},\big[\Lambda_{1},\Lambda_{2}\big]_{(\beta,\gamma)}
\Big)\Big{]}^\prime.
$

\section{Conclusion}

In physical descriptions,
what matters is the theory existing on the world-sheet,
where space-time influencing the dynamics is projected to it,
and all relevant quantities are configuration or phase space functions.
In addition symmetries are laws how background can change to 
remain in an allowed theory configuration. Differentials are standard.

In mathematical descriptions,
all variables are precisely classified and mappings between them established,
together with multi-operations resulting in field of numbers they are defined over.
While Lie derivative of the function is its time derivative,
the Lie bracket is related to scalar product of vector fields trough
self-consistency relations,
and as for forms an additional natural parring operation is relevant.
Sometimes being related to Poisson structure on a manifold,
as in Koszul bracket.

The most important for string theory turned out to be the Courant bracket,
because as shown in \cite{BDPR} the Courant algebroid represents the
suitable structure for a description of all string background 
fluxes.
There is a couple of ways to define Courant algebroid,
and two brackets defining it, both Courant and Dorfman brackets,
Courant's being antisymmetric.
Brackets are different but Dorfman bracket, given by
 $[v+\alpha,w+\beta]=[v,w]+{\cal L}_{v}\beta-i_{w}d\alpha$
sometimes goes under Courant's name \cite{DW} or under name Courant/Dorfman bracket as in \cite{Lean}.
On the other hand  Courant bracket is 
 $$[X_1 +\xi_1,X_2 +\xi_2] = [X_1,X_2]+L_{X_1}\xi_2-L_{X_2}\xi_1+
d\Big(\frac{1}{2}\big(\xi_1(X_2)-\xi_2(X_1)\big)\Big).$$
In paper \cite{HULL}, author considers the $\kappa$-bracket,
underlying the ambiguity provided by the exact term in a gauge transformation
$[A+\alpha,B+\beta]_\kappa=
 [A,B]+{\cal L}_{A}\beta-{\cal L}_{B}\alpha-
d\Big(\frac{1}{2}\kappa\big(i_{A}\beta-i_{B}\alpha\big)\Big).$
In the framework these brackets were found, the symmetrization,
which is not a naturally imposed law, influences the Leibniz rule and Leibniz identity,
making them true or making other rules present themselves,
where the above rules are calculated for the derivatives related to brackets.

In this paper we considered the C-bracket.
It is the generalization of Courant bracket in doubled space
$$[\Sigma_{1},\Sigma_{2}]=\Sigma^{N}_{[1}\partial_{N}\Sigma_{2]}^{M}-\frac{1}{2}\eta^{MN}\eta_{PQ}\Sigma^{P}_{[1}\partial_{N}\Sigma_{2]}^{Q}.$$
In double theory,
the symmetrization is related to the forms of anomaly, and consequently ways it can be cancelled. In this paper the Jacobiator is found for the 3--bracket, defined in a same manner as the 2-bracket, as the law of parameter couplings in Poisson algebra of the corresponding generators. The same as in single theory,
two brackets can be defined, although taking into consideration the indistinctness of string parameters in the anomaly,
all brackets can be represented as a definite linear combinations of a
Lie bracket components in both single and double theory.

To conclude,
geometric definitions of brackets are mainly focused on ways the objects of  dual algebras may be connected,
with the principal law being the linearity of the exterior derivative.
Physical definitions originate from evolution of physical quantities, but are set for unphysical parameters.
Here the main focus is on non-locality (dependence on the nearby points), and the ways it can influence the latter, or fade away.


\end{document}